\documentclass[useAMS,usenatbib]{mn2e}

\title[The last breath of the young GPS spectrum 
radio source PKS\,1518+047]{The last breath of the young gigahertz-peaked
spectrum radio source PKS\,1518+047}
\author[M. Orienti et al.]
  {M. Orienti$^{1,2}$\thanks{E-mail: orienti@ira.inaf.it},
M. Murgia$^{2,3}$, D. Dallacasa$^{4,2}$\\
$^1$Instituto de Astrof\'{i}sica de Canarias, c/ V\'{i}a L\'actea s/n,
E-38205  La Laguna (Tenerife), Spain\\
$^{2}$ INAF -- Istituto di Radioastronomia, Via P. Gobetti 101, I-40129 Bologna, Italy\\
$^{3}$ INAF -- Osservatorio Astronomico di Cagliari, Poggio dei Pini, Strada
54, 09012 Capoterra (CA), Italy \\
$^{4}$ Dipartimento di Astronomia, Universit\`a di Bologna, via Ranzani 1,
I-40127, Bologna, Italy }

\date{Received \today; accepted ?}

\pagerange{\pageref{firstpage}--\pageref{lastpage}} \pubyear{2002}

\def\LaTeX{L\kern-.36em\raise.3ex\hbox{a}\kern-.15em
    T\kern-.1667em\lower.7ex\hbox{E}\kern-.125emX}

\begin{document}

\label{firstpage}

\maketitle

\begin{abstract}
We present the results from multi-frequency VLBA observations from 327 MHz
to 8.4 GHz of the gigahertz-peaked spectrum 
radio source PKS\,1518+047 (4C\,04.51) aimed at studying the
spectral index distribution across the source. Further multi-frequency
archival VLA data were analysed to constrain the spectral shape of the
whole source. 
The pc-scale resolution provided by the VLBA data allows us to resolve the
  source structure in several sub-components. The analysis of their
  synchrotron spectra showed that the source components have
  steep spectral indices, suggesting that no supply/re-acceleration  
  of fresh particles is currently 
  taking place in any region of the source. By assuming the
    equipartition magnetic field of 4 mG, we found that only
  electrons with $\gamma \leq 600$, are
  still contributing to the radio spectrum, while electrons with
  higher energies have been almost completed depleted. The source
  radiative lifetime we derived is 2700$\pm$600 years.
  Considering the best fit to the overall spectrum, we find that the 
    time in which the nucleus has not been active represents
  almost 20\% of the whole source lifetime, indicating that the
  source was 2150$\pm$500 years old when the radio emission
  switched off.   
\end{abstract}


\section{Introduction}

The radio emission of extragalactic powerful radio sources is due to
synchrotron radiation from relativistic particles with a power-law
energy distribution.
They are produced in the ``central engine'', namely the active
  galactic nucleus (AGN), and reaccelerated in the hot spot,
  that is the region where the particles, channelled through the jets,
  interact with the external medium. 
The energy distribution of the
  plasma is described by a power-law: $N(E) = N_o E^{-\delta}$ 
that results in a power-law radio spectrum 
$S_{\nu}\propto  \nu^{- \alpha}$,
with a moderate steepening $S_{\nu}\propto \nu^{- \alpha + 0.5}$ 
caused by energy losses
in the form of radio emission.
The spectral steepening occurs at 
the break frequency, $\nu_{\rm b}$, which is related to the age of
relativistic electrons \citep[see e.g.][]{pacho70}. This
model, known as the {\it continuous injection} (CI) model, requires a
continuous injection of power-law distributed electrons into a
volume permeated by a constant magnetic field. The model has
been applied to the interpretation of the total spectra of the radio
sources \citep[see e.g.][]{mm99}, assuming that the
emission of the lobes (supposed to have a constant and uniform
magnetic field) dominates over that of core, jets and/or hot spots.\\ 
It is nowadays clear that powerful radio sources represent 
a small fraction of the population of the active galactic nuclei 
associated with elliptical galaxies, implying  
that the radio activity is a transient phase in the life of these
systems. 
The onset of radio activity is currently thought to be related to
merger or accretion events occurring in the host galaxy. \\
The evolutionary stage of each powerful radio source is indicated 
by its
linear size. Intrinsically compact radio sources with linear size
LS $\leq$ 1 kpc, are therefore interpreted as young radio
objects in which the radio emission originated a few thousand years
ago \cite[e.g.][]{cf95, sn00}. 
These sources are characterised by a rising synchrotron spectrum
peaking at frequencies around a few gigahertz and are 
known as gigahertz-peaked
spectrum (GPS) objects.  
When imaged with high resolution, they often display a two-sided
morphology which looks like a scaled-down version of the classical
and large (up to a few Mpc) double radio sources \citep{pm82}. 
For these sources, kinematic \citep{pc03} and
radiative \citep{mm03} studies provided ages of the order of 
10$^{3}$ and 10$^{4}$ years, supporting the idea that they are young
radio sources whose fate is probably to become classical
doubles. However, it has been claimed \citep{alexander00, marecki06} 
that a fraction of
young radio sources would die in an early stage, never becoming
the classical extended radio
galaxies with linear sizes of a few hundred kpc and ages of the order
of 10$^7$ -- 10$^8$ yr. Support for this scenario comes from a
statistical study of GPS sources by \citet{gugliu05}:
they showed that 
the age distribution of the small radio sources they
considered has a peak around 500 years. \\
So far, only a few objects have been recognised as dying radio
sources. They are difficult to find owing to their extremely
steep radio spectrum which makes them under-represented in
flux-limited catalogues. The objects 0809+404 \citep{kunert06} and
1542+323 \citep{kunert05} are possible examples of young radio sources
that are fading.
These sources, with an estimated age of 10$^{4}$ -- 10$^{5}$ years, have
been suggested as ``faders'' due to the lack of active structures,
like cores and hot-spots, although their optically-thin radio spectra
do not display the typical form expected for a fader.\\

In this paper, we present results from multi-frequency VLBA and VLA
data of the GPS radio source PKS\,1518+047 
(RA=15$^{h}$ 21$^{m}$
14$^{s}$ and DEC= 04$^{\circ}$ 30$^{'}$ 21$^{''}$, J2000), identified with a
quasar at redshift $z = 1.296$ \citep{sk96}. 
It is a powerful radio
source (Log $P_{\rm 1.4\, GHz}\,\, {\rm (W/Hz)} = 28.53$),
with a linear size of 1.28 kpc.
This source was selected from the GPS sample of \citet{cstan98} 
on the basis of its steep spectrum $\alpha_{\rm
  1.4 GHz}^{\rm 8.4 GHz} = 1.2$, uncommon in young radio
sources. The goal of this paper is to 
determine by means of the analysis of the 
radio spectrum and its slope in the optically thin region,
whether this source is in a phase where its radio activity
has, perhaps temporarily, switched off.\\

Throughout this paper, we assume the following cosmology: $H_{0} = 71\,
{\rm km\, s^{-1}\, Mpc^{-1}}$, $\Omega_{\rm M} = 0.27$ and
$\Omega_{\Lambda} = 0.73$, in a flat Universe. At the redshift of the
target 1$^{''}$ = 8.436 kpc. The spectral index
is defined as 
$S {\rm (\nu)} \propto \nu^{- \alpha}$.\\

\section{Radio data}

The target source was observed on March 5, 2008 (project code
BD129) with the VLBA at 0.327, 0.611 (P band),
1.4 and 1.6 GHz (L band) with a recording band width of 16 MHz at 256
Mbps for 20 min in L band and 1 hr in P band. 
The correlation was performed at the VLBA correlator in Socorro.
These observations have been complemented with archival VLBA data at 4.98
(C band) and 8.4 GHz (X band) carried out on March 28, 2001 (project
code BS085).
The data reduction was carried out with the NRAO AIPS package. \\
During the observations, the gain corrections
were found to have variations  
within 5\% in L, C and X bands, and 10\% in P band respectively. In P
band, the antennas located in Kitt Peak, Mauna Kea and North Liberty
had erratic system temperatures, probably due to local RFI, and
their data had to be flagged out completely. The resulting loss
of both resolution and sensitivity did not affect the source structure
and flux density.\\
To constrain the spectral shape of the whole source, we analysed
archival VLA data at 0.317, 1.365, 1.665, 4.535, 4.985, 8.085, 8.465, 14.940
and 22.460 GHz, obtained on August 22, 1998 (project code
AS637). 
The data reduction was carried out with the standard
procedure implemented in the NRAO AIPS package. Uncertainties in the
determination of the flux density are dominated by amplitude
calibration errors, which are found to be around 3\% at all
frequencies.\\
VLA and VLBA final images were obtained after a number of phase-only
self-calibration iterations. 
At 4.9 and 8.4 GHz, besides the {\it full resolution} VLBA image, we
also produced a {\it lower resolution} image using the same {\it
uv}-range, image sampling and restoring beam of the 1.6 GHz, in order
to produce spectral index maps of the source. Images
at the various frequencies were aligned using the task LGEOM by
comparing the position of the compact and bright components at each
frequency.\\
Flux density and
deconvolved angular sizes 
were measured by means of the task JMFIT which performs a
Gaussian fit to the source components on the image plane, 
or, in the case of extended components, by TVSTAT which performs
an aperture integration on a selected region on the image plane. \\
The parameters derived in this way 
are reported in Table \ref{alfa}. 
About 8\% and 20\% of the total flux density is missing in our VLBA
images at 4.9 and 8.4 GHz respectively 
if we consider the VLA measurements. 
It is likely that such a missing flux density is
from a steep-spectrum and diffuse emission which is not sampled by
the VLBA data due to the lack of appropriate short spacings. \\ 

\begin{table}
\begin{center}
\begin{tabular}{|c|c|c|c|}
\hline
Freq.&Tot$^{a}$&N$^{b}$&S$^{b}$\\
GHz&mJy&mJy&mJy\\
\hline
&&&\\
0.32&2278$\pm$38&1136$\pm$117&512$\pm$58\\
0.61&4050$\pm$200$^{c}$&1778$\pm$182&897$\pm$98\\
0.96&4720$\pm$240$^{d}$& - & - \\
1.4 &3937$\pm$120&2468$\pm$123&1555$\pm$78\\
1.6 &3376$\pm$101&2008$\pm$101&1311$\pm$56\\
3.9 &1460$\pm$50$^{d}$& - & - \\
4.5 &1164$\pm$30& - & - \\
4.9 &1013$\pm$30&392$\pm$19&479$\pm$24\\
5.0 &994$\pm$52$^{e}$& - & - \\
8.1 &486$\pm$15& - & - \\
8.4 &441$\pm$13&148$\pm$7&232$\pm$11\\
11.1&305$\pm$30$^{d}$& - & - \\
15  &165$\pm$1& - & - \\
22  &80$\pm$5& - & - \\
&&&\\
\hline
\end{tabular}
\caption{Multi-frequency flux density of PKS\,1518+047 
and of its northern and
  southern components. $a$ = archival VLA data; $b$ = VLBA data; $c$ = WSRT data from
  \citet{cstan98}; $d$ = RATAN data from
  \citet{cstan98}; $e$ = WSRT data from \citet{xiang06}.  }
\label{alfa}
\end{center}
\end{table}

\section{Results}

\subsection{Source structure}

\begin{figure*}
\begin{center}
\includegraphics{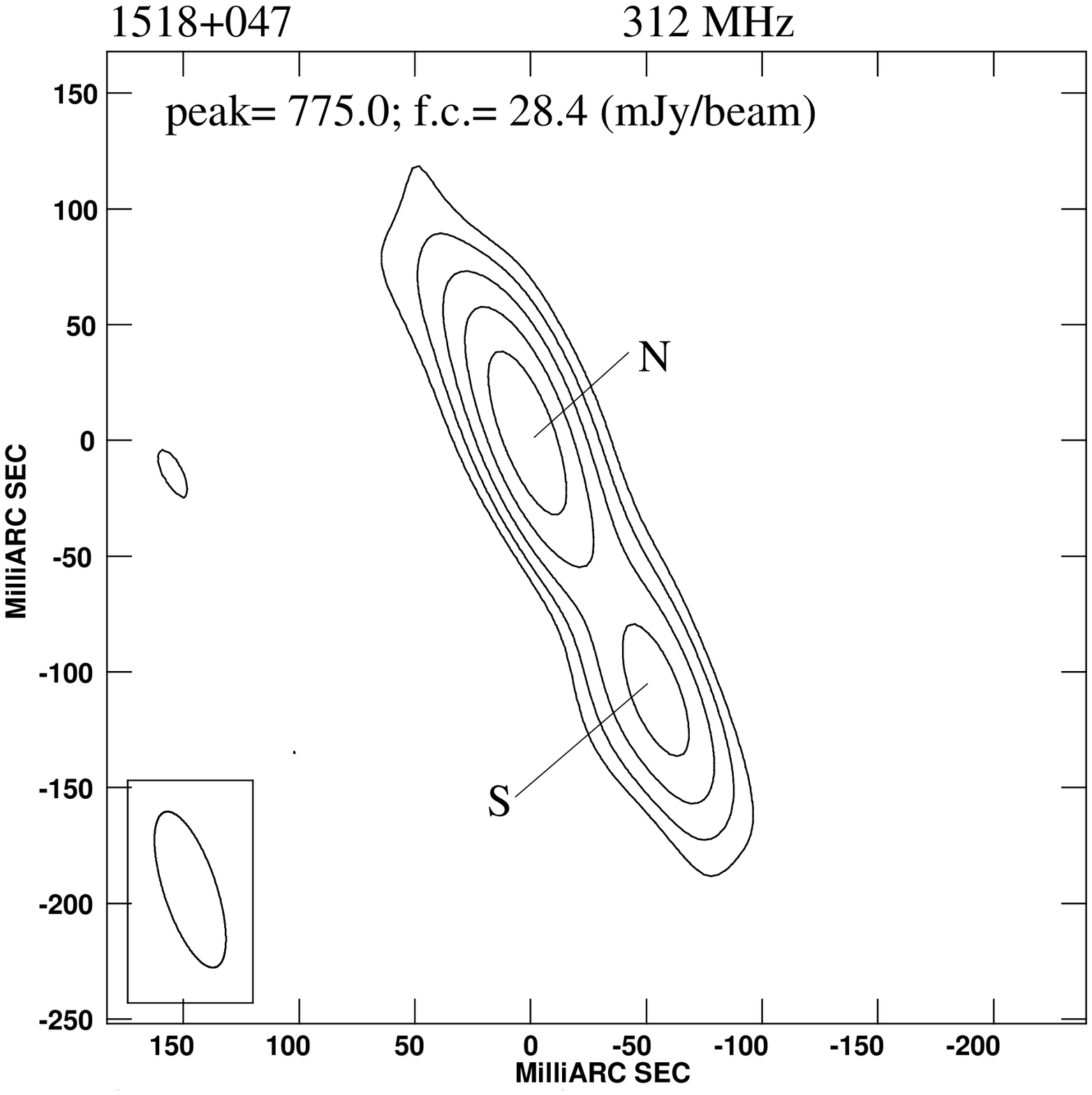}
\includegraphics{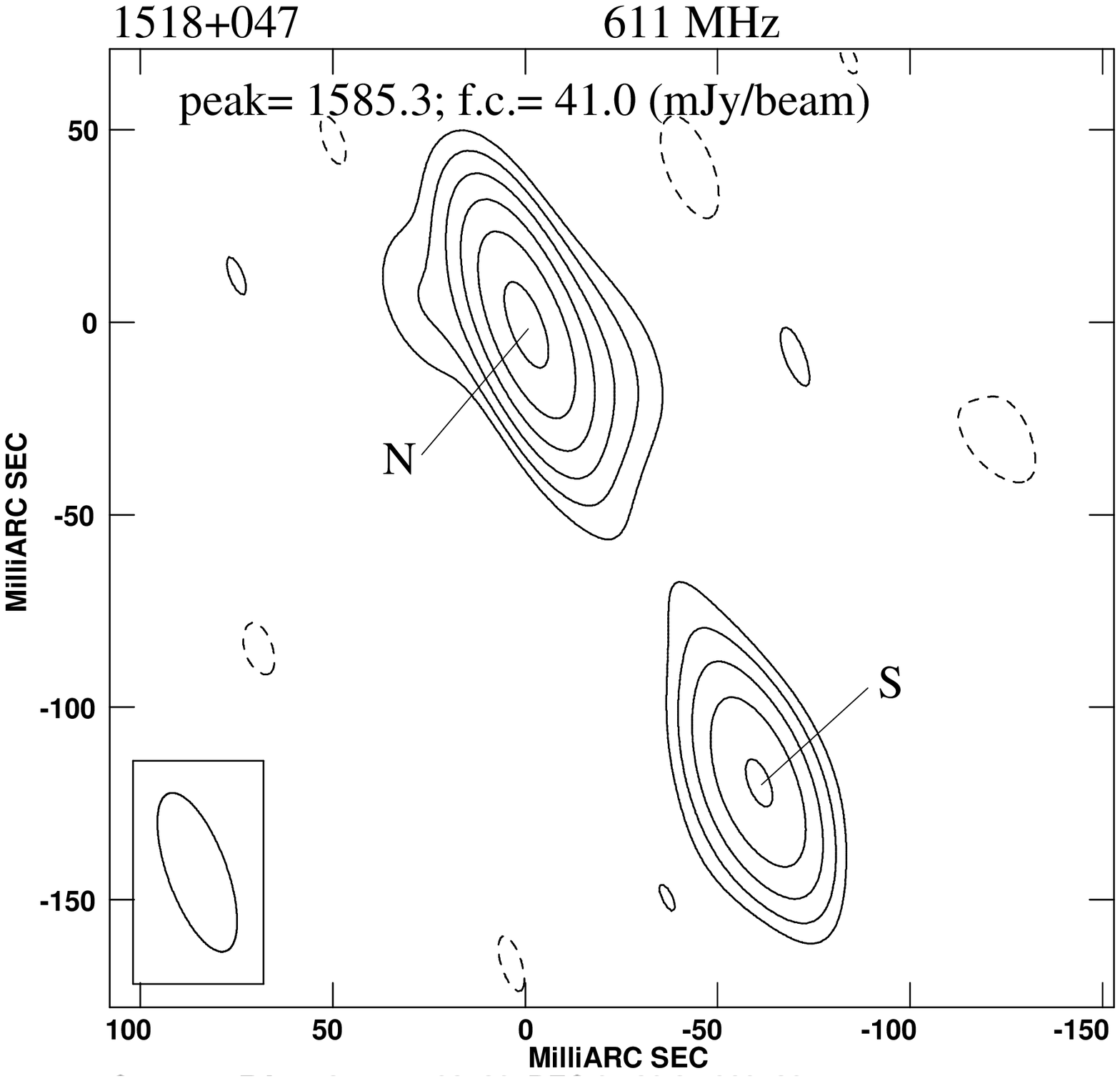}
\includegraphics{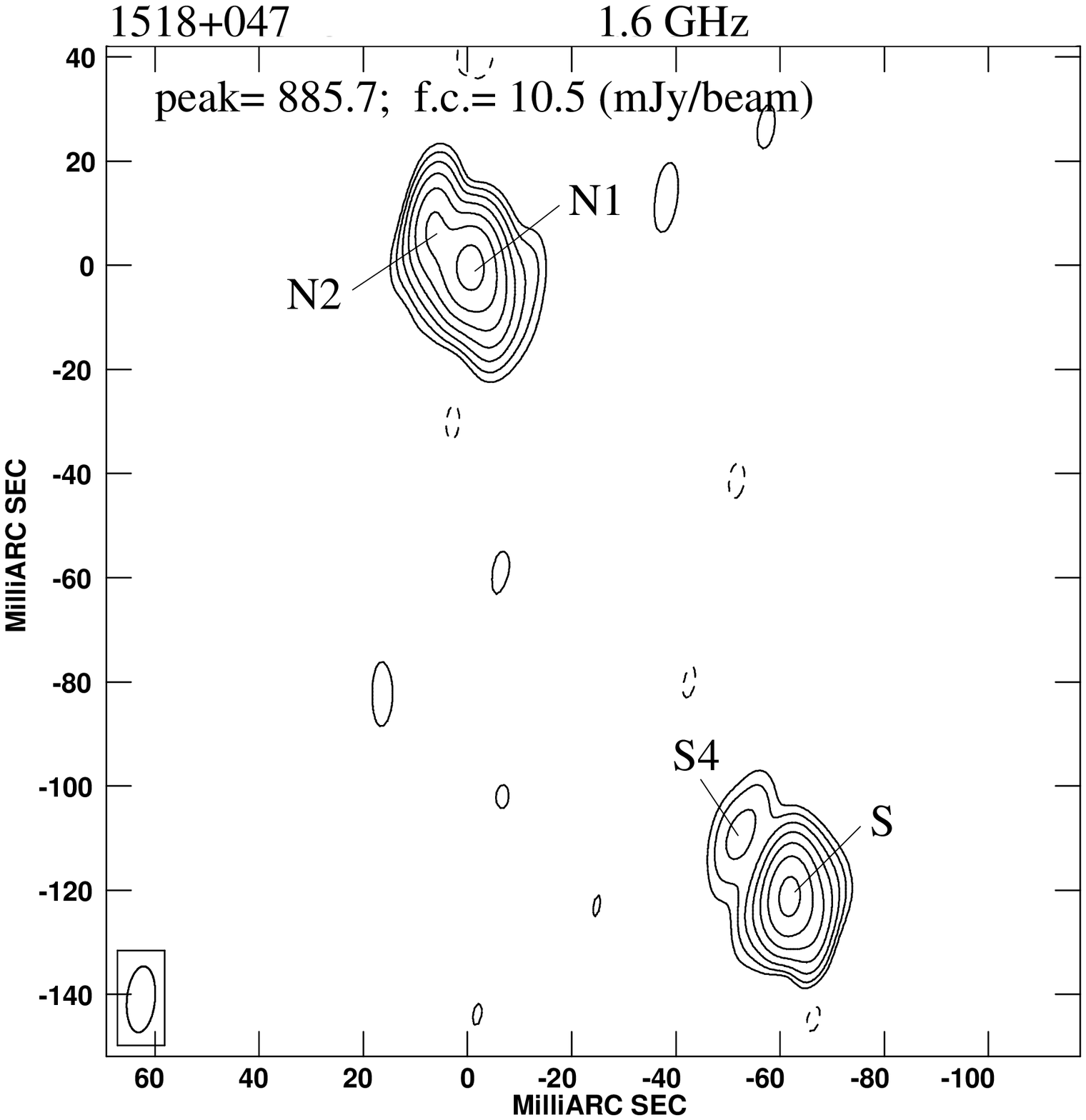}
\includegraphics{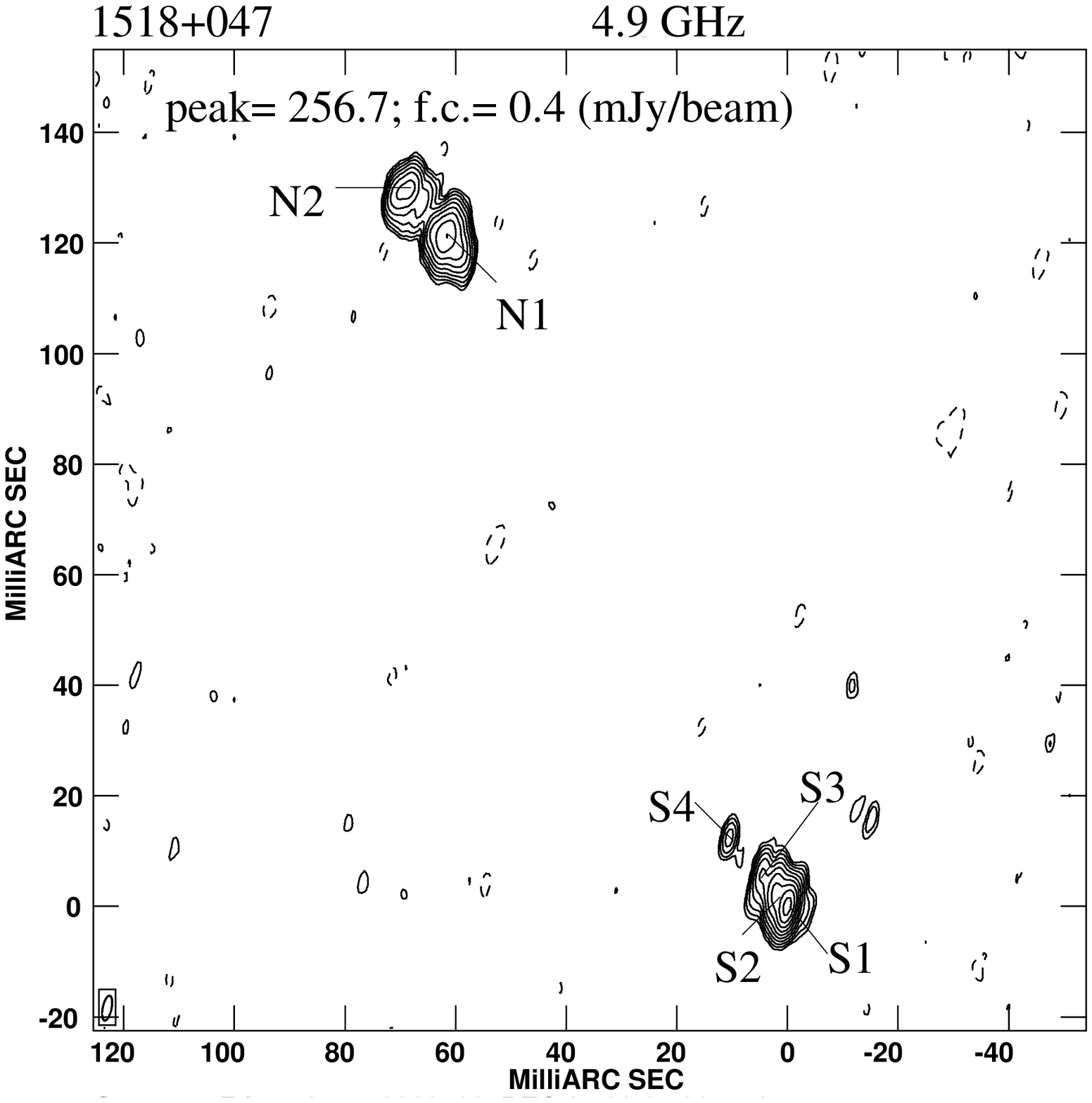}
\includegraphics{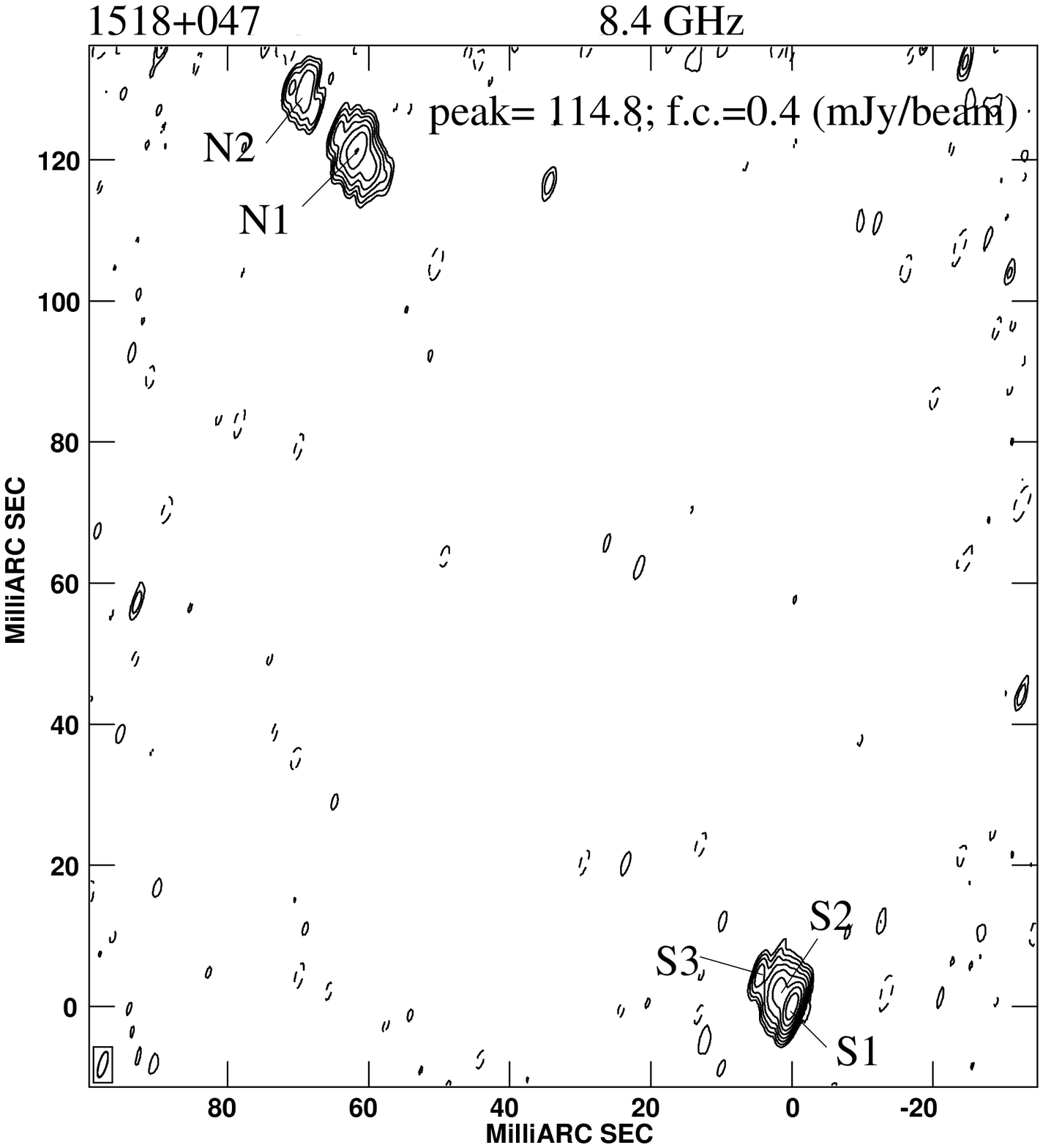}
\vspace{13cm}
\caption{VLBA images at 0.327, 0.611, 1.6, 4.9 and 8.4 GHz of
  PKS\,1518+047. On the images we show the observing frequency, the peak
  flux density and the first contour intensity ({\rm f.c.}) which is 3
  times the 1$\sigma$ noise level measured on the image
  plane. Contours increase of a factor 2. The beam is plotted on the
  bottom left corner.} 
\label{images}
\end{center}
\end{figure*}

\begin{figure*}
\begin{center}
\includegraphics{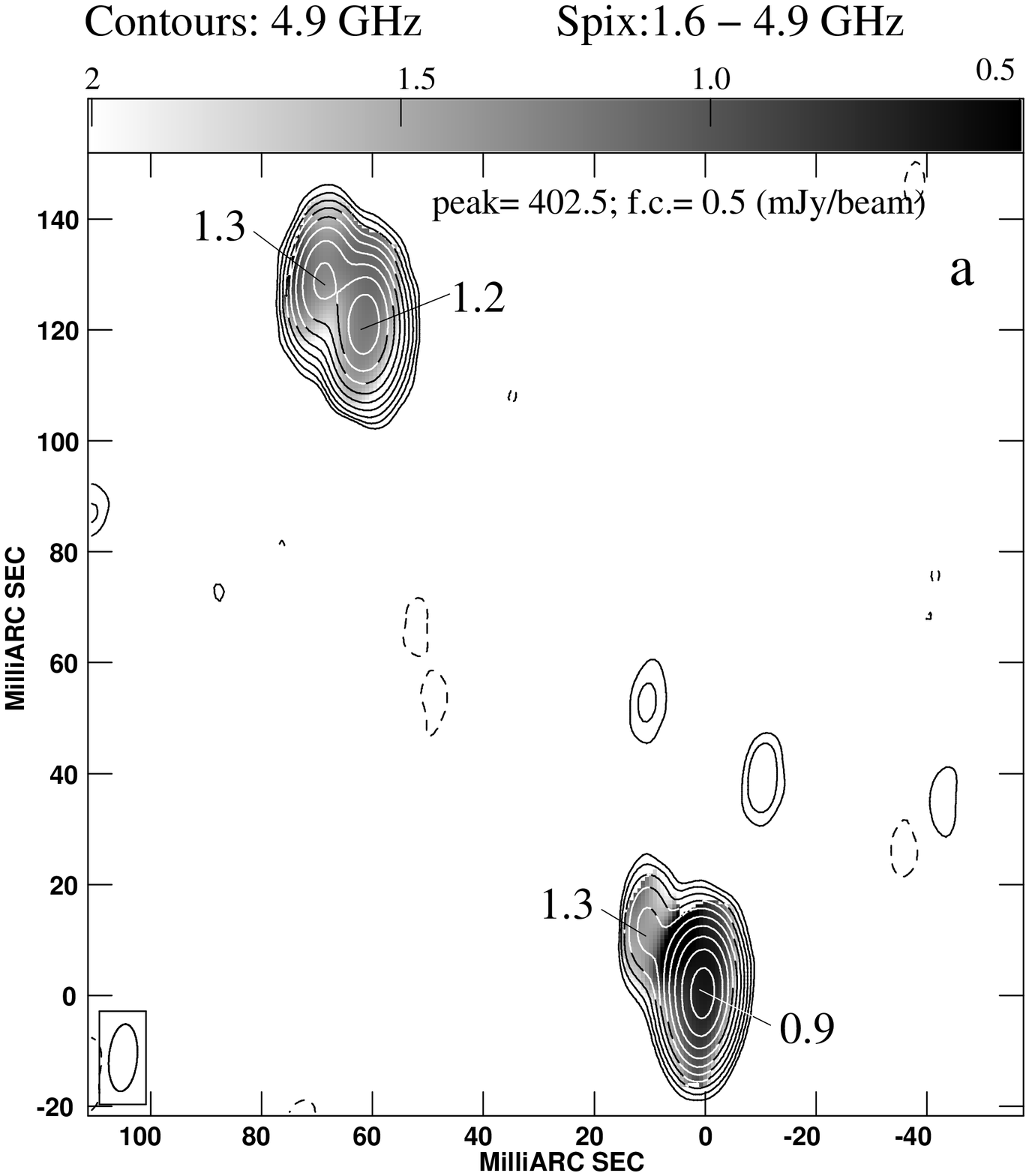}
\includegraphics{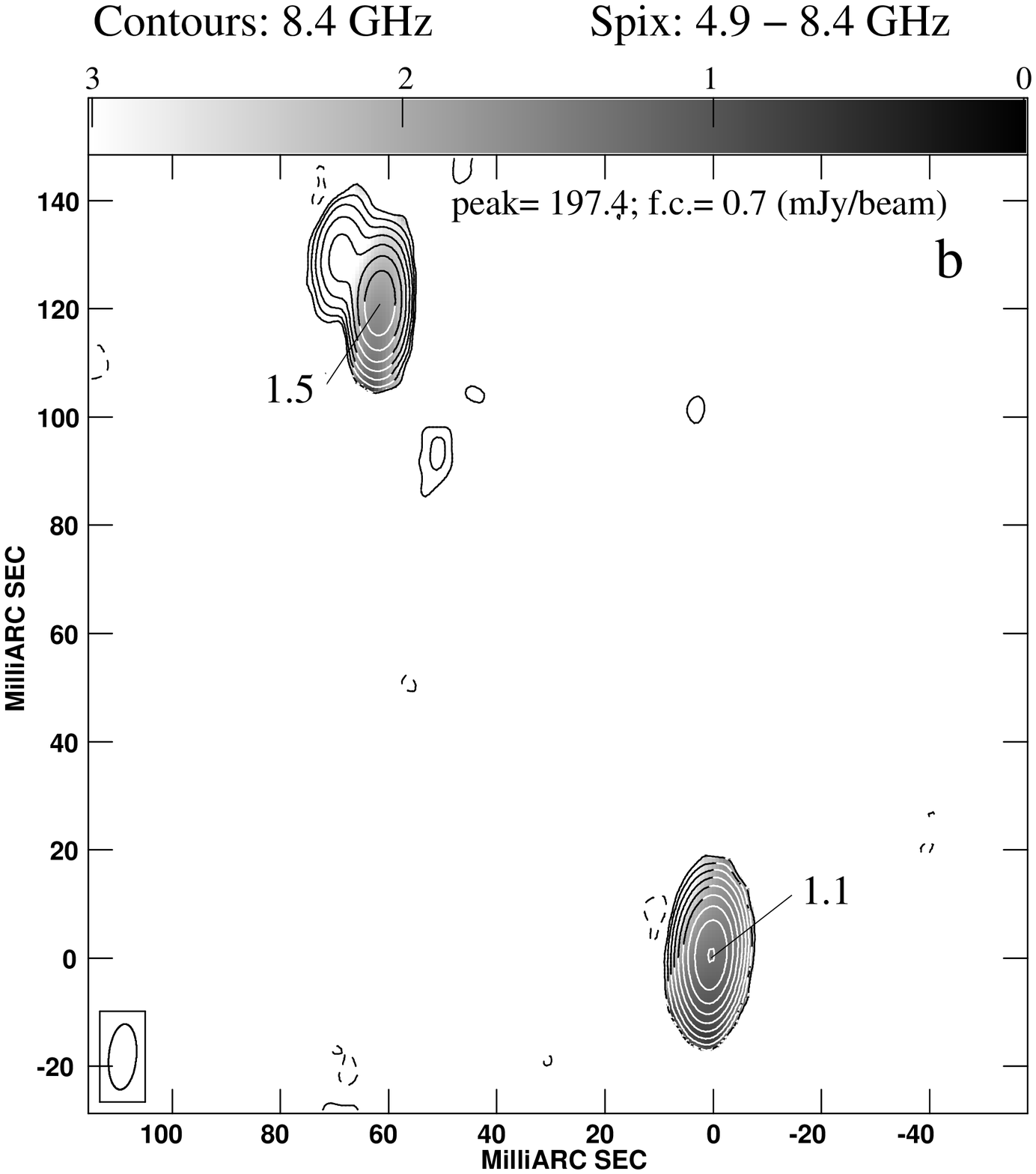}
\includegraphics{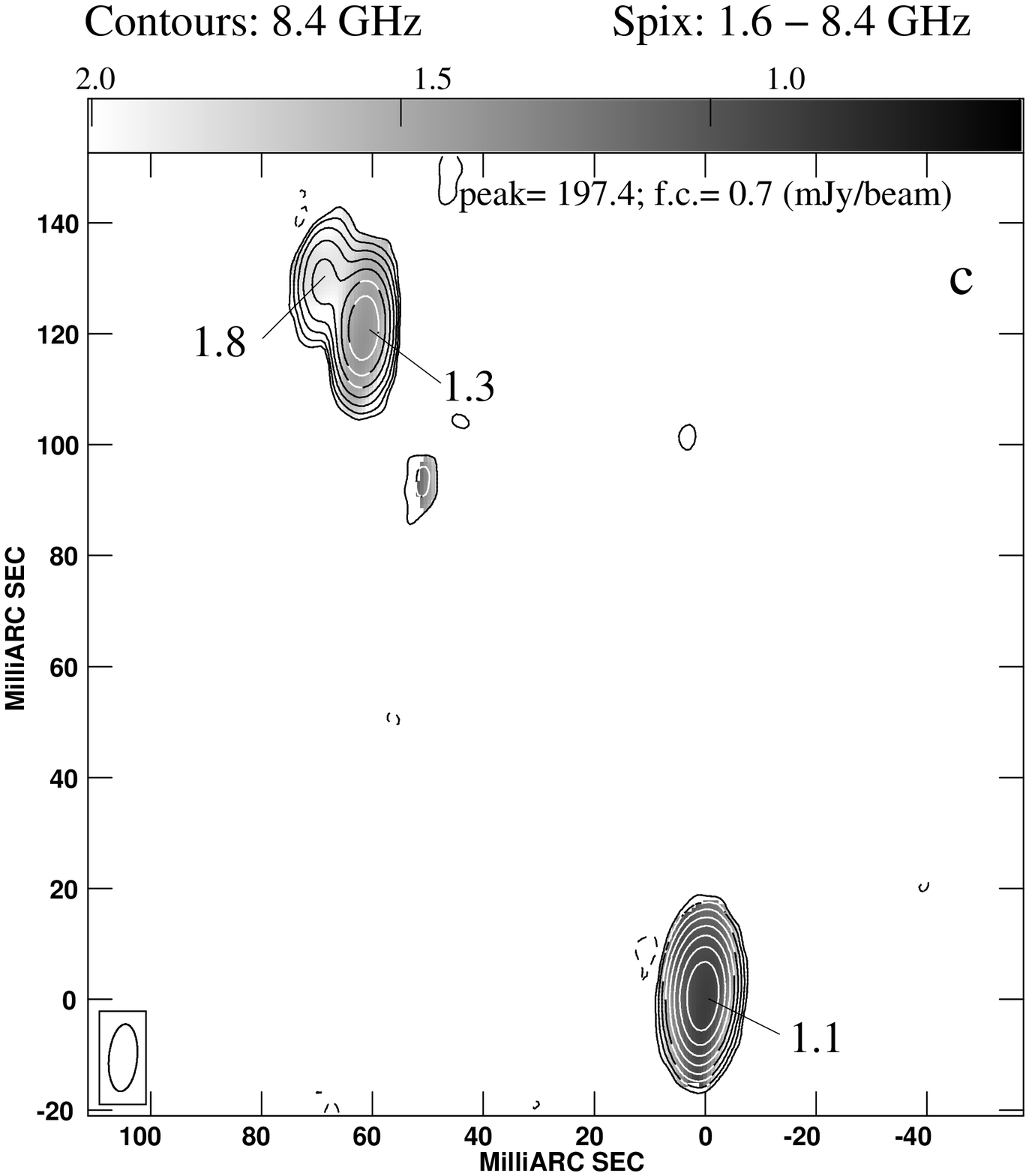}
\vspace{8cm}
\caption{VLBA spectral index images of PKS\,1518+047 
between 1.6 and 4.9 GHz ({\it panel
a}), between 4.9 and 8.4 GHz ({\it panel b}), and between 1.6 and 8.4
GHz ({\it panel c}). On the images we show the observing frequency 
of the contours, the peak
  flux density and the first contour intensity ({\rm f.c.}) which is 3
  times the 1$\sigma$ noise level measured on the image
  plane. Contours increase of a factor 2. The beam is plotted on the
  bottom left corner. The grey scale is shown by the wedge at the top
  of each image.} 
\label{spix}
\end{center}
\end{figure*}

Multi-frequency VLBA observations with parsec-scale resolution allow us to
resolve the structure of PKS\,1518+047 into two main source components
(Fig. \ref{images}),
separated by 135 mas (1.1 kpc), in agreement with
previous images by \citet{dd98},
\citet{xiang02}, and \citet{xiang06}. Both the
northern and southern components have angular size of 23$\times$15 mas$^{2}$.\\
At the highest frequencies, where we achieve the best resolution, the
northern component is resolved into two sub-structures (labelled 
N1 and N2) separated by 11 mas (90 pc) and position angle of
11$^{\circ}$. 
N1 accounts for 295 and 119 mJy at 4.9 and 8.4
GHz, respectively, with angular size of 7.7$\times$6.8 mas$^{2}$. 
Component N2 is fainter than N1 and it accounts for
97 and 29 mJy at 4.9 and 8.4 GHz, respectively, with angular size of  
7.4 $\times$5 mas$^{2}$.   
The southern complex is resolved into 4 compact regions located at
3, 7 and 16 mas (25, 60 and 130 pc for S2, S3 and S4 respectively)
with respect to S1, all with almost the same position angle of
40$^{\circ}$. 
The resolution, not adequate to properly fit the several components of the
southern lobe, does not
allow us to reliably determine the observational parameters of each single
sub-component.  \\
The spectral index in both the northern and southern complexes,
computed considering the integrated component flux density, is
steep, with $\alpha_{1.4}^{8.4} = 1.5\pm0.1$ and $= 1.0\pm0.1$,
respectively. Errors on the spectral indices were calculated
following the error propagation theory.
The analysis of the spectral index distribution in the southern
complex (Fig. \ref{spix}) shows
a steepening of the spectral index going from S1 inwards to  
S4, suggesting that S1 is likely the last place
where relativistic electrons coming from the source
core were re-accelerated. 
In the northern lobe, the spectral index between 1.6 and 4.9 GHz is almost
constant across the component, showing a small gradient going from N1
towards N2, while in the spectral index maps at higher frequencies it
steepens from N1 to N2 (Fig. \ref{spix}).
The steep spectral indices showed by the
northern and southern components suggest that no current particle
acceleration across the source is taking place, indicating that  
active regions, like conventional jet-knots and hot spots, are
no longer present. \\
At 327 and 611 MHz, i.e. in the optically-thick part of the spectra, 
the low spatial resolution does not allow us to separate the individual
sub-components. However, their spectral indices, obtained
considering the integrated flux density, are
$\alpha^{611}_{327} \sim$ -0.6$\pm$0.2 and -0.9$\pm$0.2
for the northern and southern
component respectively, that is much flatter than the value expected in
the presence of ``classical'' synchrotron self-absorption from a
homogeneous component, indicating that what we see is the
superposition of the spectra of  
the various sub-components, each characterised by its own
  spectral peak occurring in a range of frequencies and then causing the
  broadening of the overall spectrum (i.e. the northern and
  southern components are far from being homogeneously filled by
  magnetised relativistic plasma). \\  
 
\subsection{The spectral shape}

To understand the physical processes taking
place in this source we fit the optically-thin part of the 
overall spectrum, as well as the
spectra of the northern and southern components, assuming two different
models. The first model assumes that
fresh relativistic particles are continuously injected in the
source ({\it CI model}), while in the second model the continuous supply of
particles is over and the radio source is already in the relic phase
({\it CI OFF model}). \\

The CI synchrotron model is described by three parameters:
\begin{itemize}
\item[i)] $\alpha_{inj}$, the injection spectral index;
\item[ii)] $\nu_{b}$, the break frequency;
\item[iii)] $norm$, the flux normalization.
\end{itemize}
 
As the injection of fresh particles stops (CI OFF model), 
a second break, $\nu_{\rm b2}$, appears at high
frequencies, and beyond that the spectrum cuts off exponentially.
This second break is related to the first according to:
 
\begin{equation}   
\nu_{\rm b2} = \nu_{\rm b} \left( \frac{t_{\rm s}}{t_{\rm OFF}} \right)^{2}  
\label{eq_br2}
\end{equation}

\noindent where $t_{\rm s} $ is the total source's age and $t_{\rm
  OFF}$ is the time elapsed since the injection switched off 
  \citep[see e.g.][]{komissarov94, slee01, parma07}.
Indeed, compared to the basic CI model, the CI OFF model 
is characterized by one more free parameter:

\begin{itemize}
\item[i)] $\alpha_{inj}$, the injection spectral index;
\item[ii)] $\nu_{b}$, the lowest break frequency;
\item[iii)] $norm$, the flux normalization;
\item[iv)]  $t_{\rm OFF}/t_{s}$, i.e. the relic to total source age ratio.
\end{itemize}

\noindent The spectral shapes of both models cannot be described 
by analytic equations
and must be computed numerically (see e.g. Murgia et al. 1999 and
Slee et al. 2001 for further details).\\

\begin{figure}
\begin{center}
\includegraphics{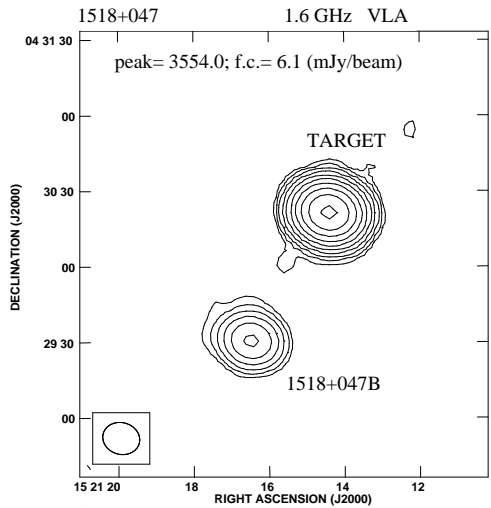}
\vspace{7cm}
\caption{VLA image at 1.6 GHz of PKS\,1518+047. Another source, 
1518+047B, 
is located at $\sim$1
  arcminute in the south-east direction from the target.}
\label{vla}
\end{center}
\end{figure}

\begin{figure*}
\begin{center}
\includegraphics{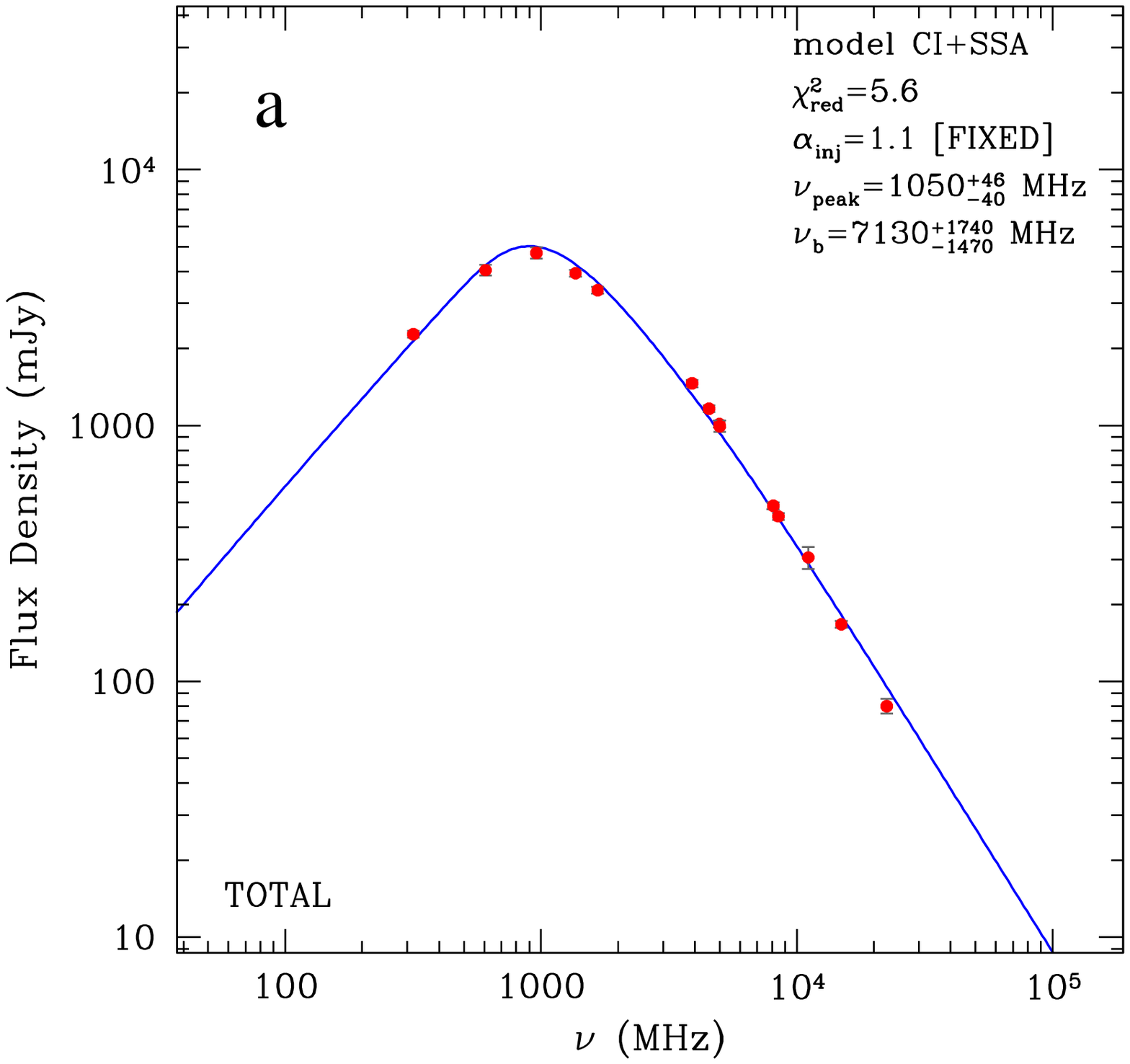}
\includegraphics{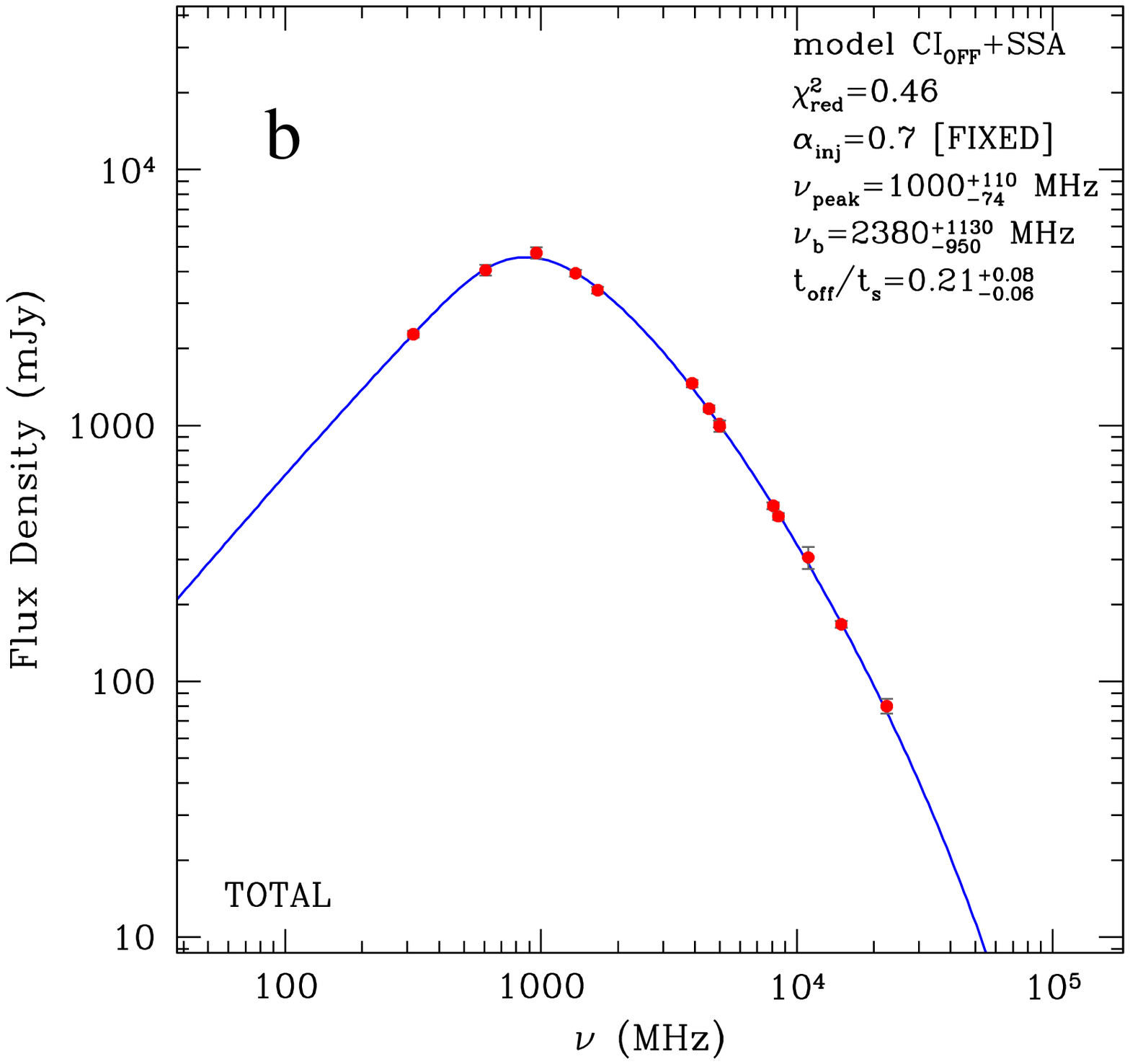}
\includegraphics{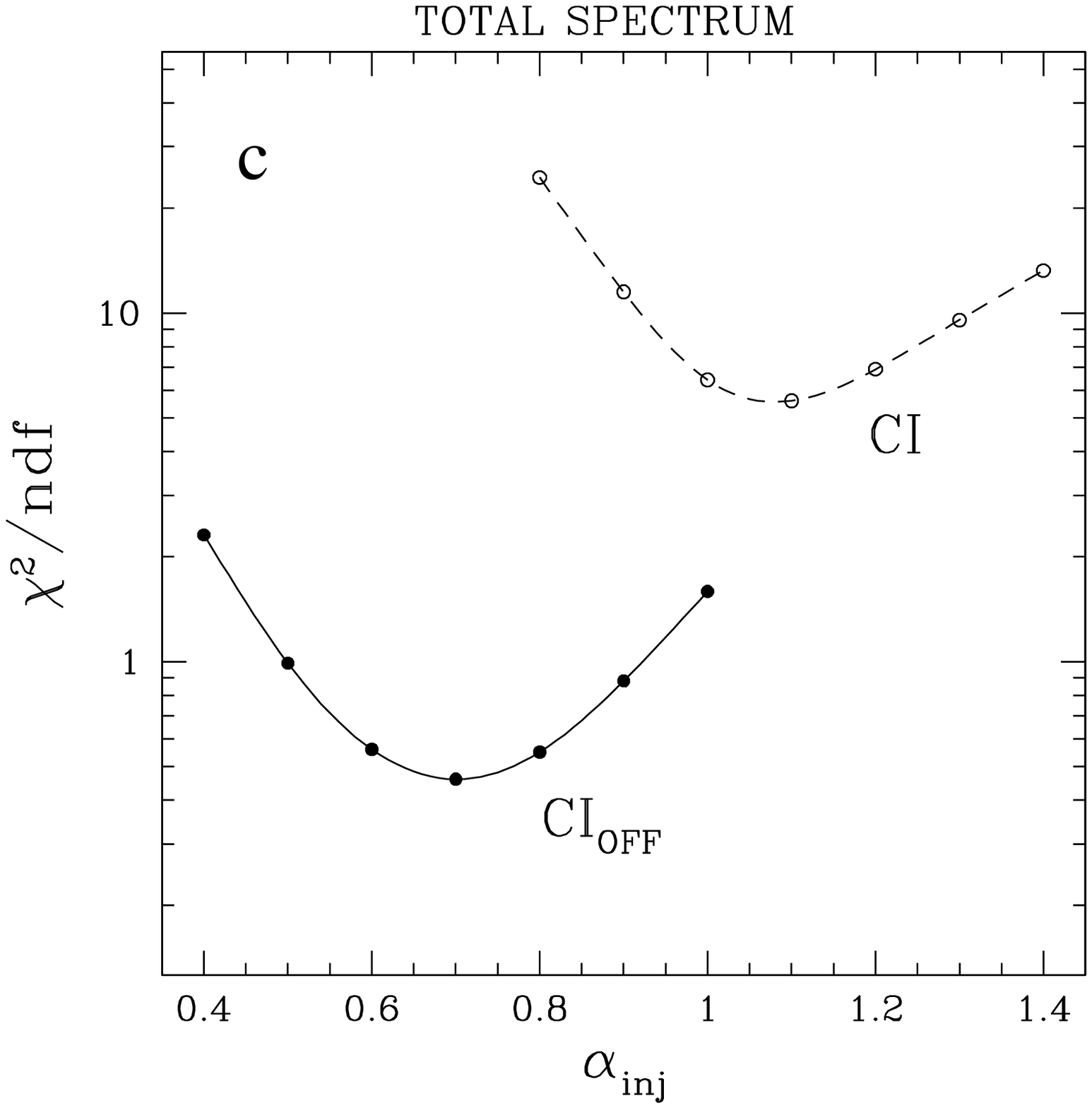}
\vspace{6cm}
\caption{The best fit to the overall spectrum of PKS\,1518+047 using the CI
  ({\it left}) and the CI OFF models ({\it center}), and the reduced
  chi-square versus the injection spectral index ({\it
  right}). Error bars are rather small, and often result within the
  symbol.}
\label{spettro_totale}
\end{center}
\end{figure*}

\begin{figure}
\begin{center}
\includegraphics{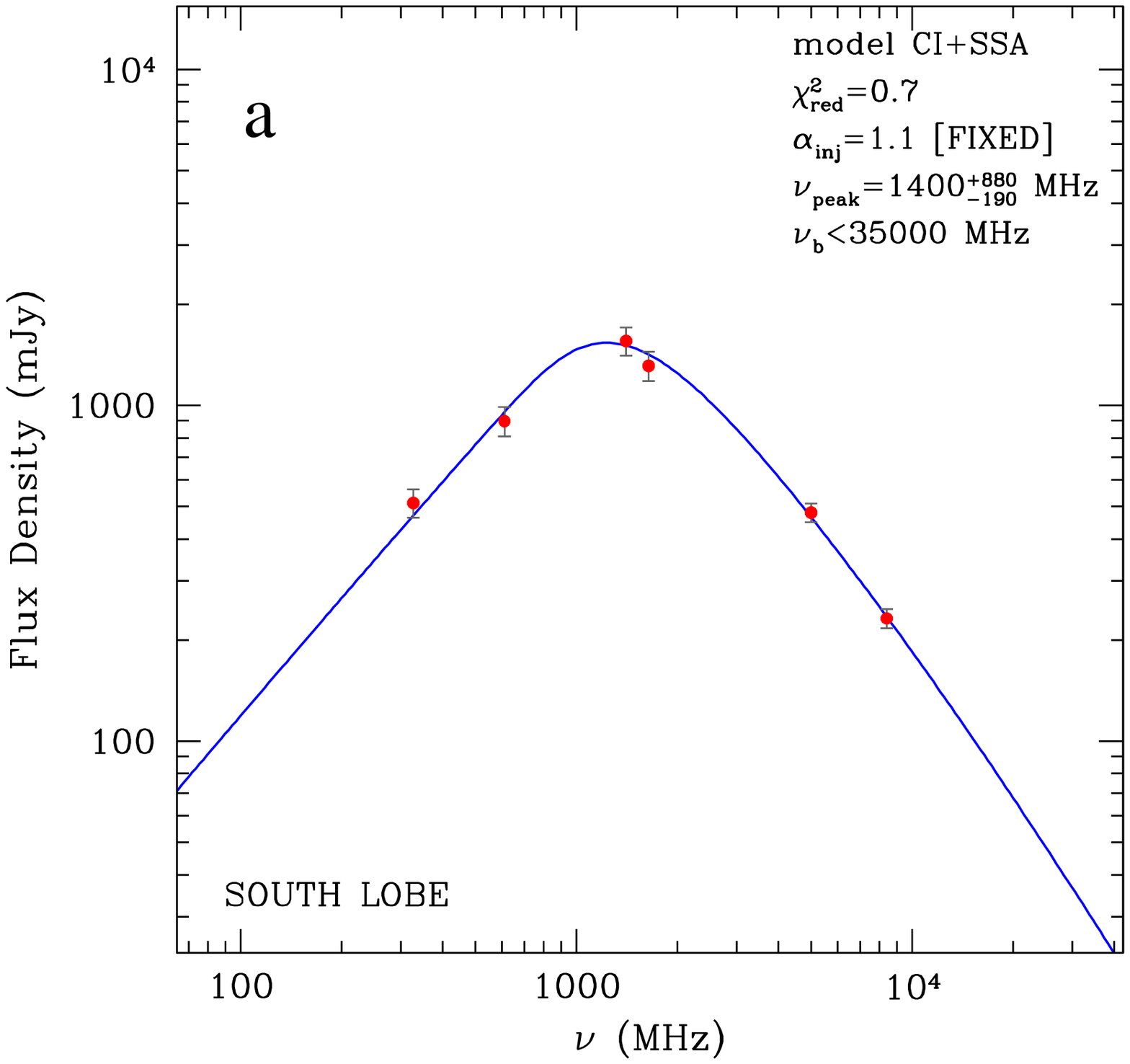}
\includegraphics{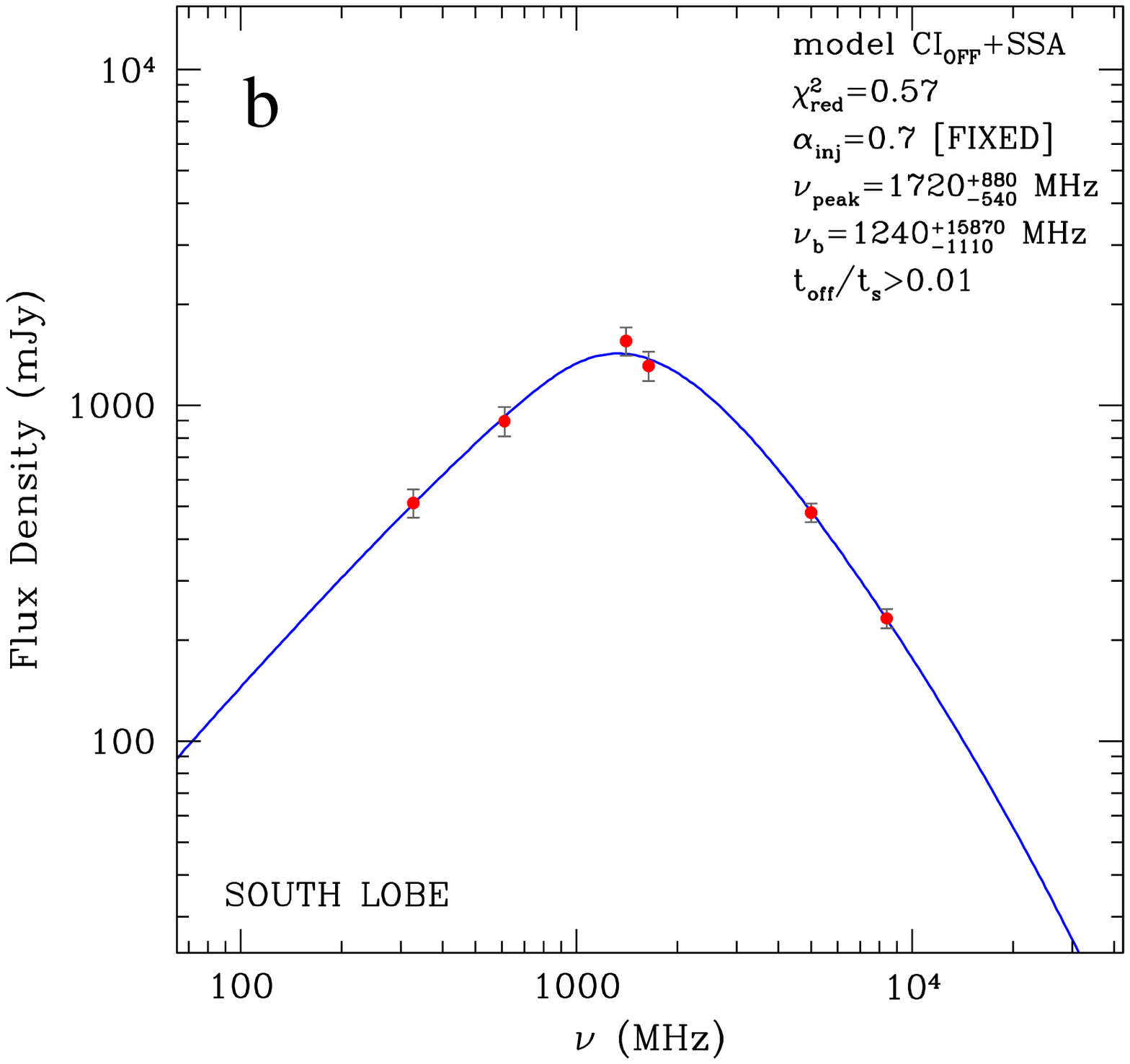}
\vspace{13.5cm}
\caption{The best fit to the spectrum of the southern component of 
PKS\,1518+047 using the CI ({\it top})
and the CI OFF models ({\it bottom}).}
\label{spettro_sud}
\end{center}
\end{figure}

\begin{figure}
\begin{center}
\includegraphics{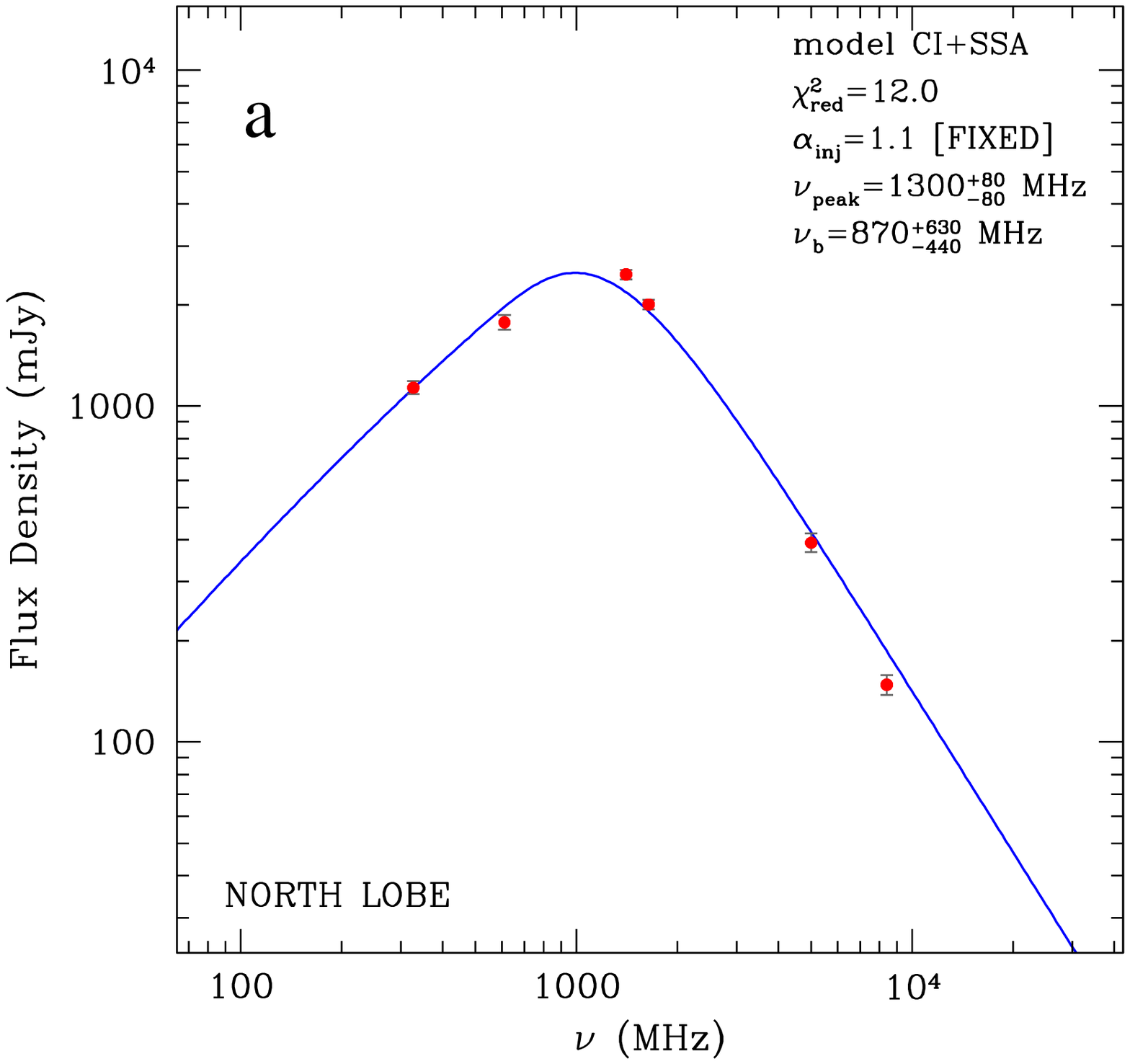}
\includegraphics{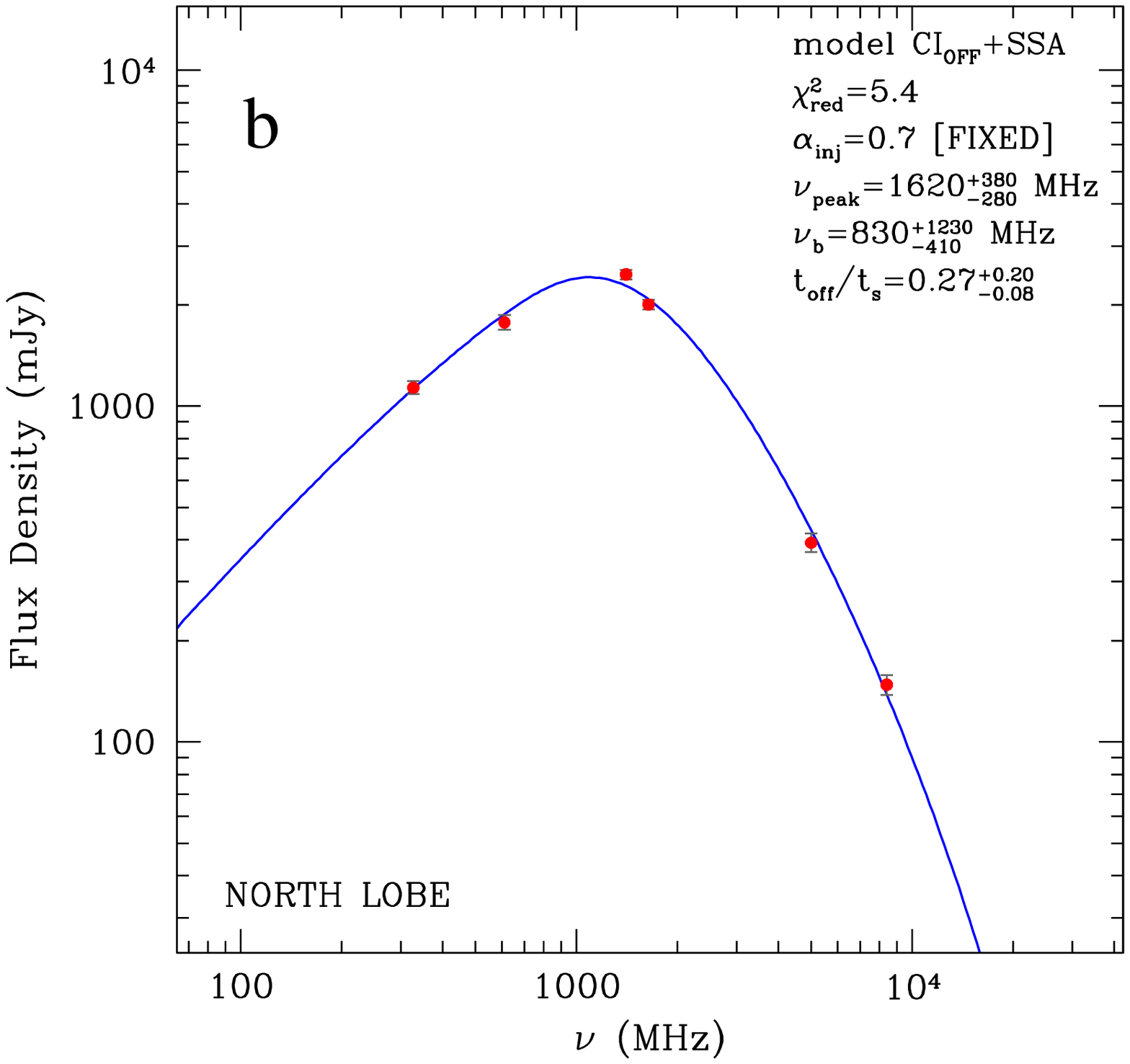}
\vspace{13.5cm}
\caption{The best fit to the spectrum of the northern component of 
PKS\,1518+047 using the CI ({\it top})
and the CI OFF models ({\it bottom}).}
\label{spettro_nord}
\end{center}
\end{figure}

In the study of the overall spectrum, in addition to the
archival VLA data analysed in this paper, we consider
also RATAN observations at 0.96, 3.9, and 11.1 GHz 
\citep{cstan98}, and WSRT observations at 5 GHz 
\citep{xiang06}, to have a better frequency sampling. Due to the
presence of another radio source located at about 1 arcminute from the
target (Fig. \ref{vla}), we do not consider observations with
resolution worse than $\sim$1$^{'}$. \\
During the fitting procedure, a particular care was taken in choosing
  the most accurate injection spectral index. The peculiar shape of
  the radio spectrum of this 
  source does not allow us to directly derive the injection
  spectral index from the optically-thin emission below
  the break, because it falls below the peak frequency where the
  spectrum is absorbed. For this reason, we fit the spectrum assuming
  various injection spectral indices, choosing the one that provides the
  best reduced chi-square (Fig. \ref{spettro_totale}c).\\  
The best fit we obtain with the CI
  model implies a very steep injection spectral index $\alpha_{\rm
  inj} = 1.1$, that is reflected in an uncommon 
electron energy distribution $\delta = 3.2$,
  (Fig. \ref{spettro_totale}a). 
On the other hand, the CI
  OFF model provides a more common injection spectral index
  $\alpha_{\rm inj} = 0.7$ (Fig. \ref{spettro_totale}b). 
Furthermore, the comparison between the
  reduced chi-square of the different models (Fig. \ref{spettro_totale}c) 
  shows that the CI OFF model is more accurate in fitting the data. In
  the context of the CI OFF model, we derive 
 a break frequency $\nu_{\rm br} =$
  2.4 GHz, and $t_{\rm OFF}/t_{s} = 0.2$. A similar result is found
for the spectrum of the southern component. As in the overall
  spectrum, both CI and CI OFF models well reproduce the spectral
  shape (Fig. \ref{spettro_sud}), 
but the former implies an uncommonly steep injection spectral index. \\
A different result is found
  in the analysis of the northern lobe. 
  In this case, we
  find that the CI OFF model with $\alpha_{\rm inj} = 0.7$ well
  reproduces the spectral shape, 
  providing a break frequency
  $\nu_{\rm br} = 0.8$ GHz and $t_{\rm OFF}/t_{s} \sim 0.27$
  (Fig. \ref{spettro_nord}b), while the CI model provides a worse
  chi-square even considering a steep injection spectral index
  (Fig. \ref{spettro_nord}a).

The optically-thick part of the spectra is well modelled by an absorbed
spectrum with $\alpha_{\rm thick} = -1.2\pm0.1$, that is different
from the canonical -2.5 expected in the presence of synchrotron
self-absorption from a homogeneous component. This result, together
with the source structure resolved in several sub-components, indicates
that in the optically-thick regime the observed spectra are the
superposition of the spectra of many components that cannot be
resolved due to resolution limitation.\\
The best fit parameters obtained with the CI and CI OFF models, and assuming
synchrotron self-absorption (SSA) are
  reported in Table \ref{best_fit}.\\

\subsection{Physical parameters}

Physical parameters for the radio source components were computed
assuming equipartition condition and using standard formulae
\citep{pacho70}. We considered 
particles with energy between $\gamma_{\rm min} =$100 and $\gamma_{\rm
  max}=$600. The high energy cut-off corresponds to the value for
which the break
frequency in the observer frame occurs at 2.4 GHz, 
as obtained from the best fit to the
model (Section 3.2).
Proton and electron energy densities are assumed
to be equal, and the spectral index is $\alpha = 0.7$ (see Section 3.2).
We assume that the volume $V$ of the emitting regions is a
prolate spheroid:

\begin{displaymath}
V = \frac{\pi}{6}\, a b^{2} \phi
\end{displaymath}

\noindent where $a$ and $b$ are the spheroid major and minor axes, 
and $\phi$ is the filling factor. 
In the case of the entire radio source, we consider
$a=$140 mas, $b=$25 mas, and $\phi = 1$.
With these assumptions, we obtain an
equipartition magnetic field $H_{\rm eq} =$ 4 mG, 
and a minimum energy
density $u_{\rm min} = 1.4 \times 10^{-6}$ erg/cm$^{3}$. 
In the case of component N and S we assume $a=23$ mas, $b=$15 mas and
$\phi$=1, and we infer a magnetic field of 7 mG,
and a minimum energy density of $=$5$\times$10$^{-6}$ erg/cm$^{3}$,
in agreement with the values derived for the entire source structure.\\

\begin{table*}
\begin{center}
\begin{tabular}{|c|c|c|c|c|c|c|}
\hline
Comp.&Model&$\alpha_{\rm inj}$&$\nu_{\rm p}$&$\nu_{\rm b}$&$t_{\rm OFF}/t_{s}$&$\chi^{2}_{\rm red}$\\
 & & &MHz&MHz& &\\
\hline
&&&&&&\\
N&CI&1.1&1300$_{-80}^{+80}$&870$_{-440}^{+630}$& - &12.0\\
N&CI OFF&0.7&1620$_{-280}^{+380}$&800$_{-410}^{+1230}$&0.27&5.4\\
S&CI&1.1&1400$_{-190}^{+880}$&$<$35000& - &0.7\\
S&CI OFF&0.7&1720$_{-540}^{+880}$&1260$_{-1110}^{+15870}$&$>$0.01&0.57\\
Tot&CI&1.1&1050$_{-40}^{+46}$&7130$_{-1470}^{+1740}$& - &5.6\\
Tot&CI OFF&0.7&1000$_{-74}^{+110}$&2380$_{-950}^{+1130}$&0.21&0.46\\
&&&&&&\\
\hline
\end{tabular}
\caption{Best fit parameters obtained with the CI and CI OFF models + 
SSA for the
  source PKS\,1518+047 and for its northern and southern components.}
\label{best_fit}
\end{center}
\end{table*}

\section{Discussion}

In general the shape of the synchrotron spectrum of active
radio sources is the result of the interplay between freshly injected
relativistic particles and energy losses that cause a depletion of
high-energy particles which  
results into a steepening of the spectrum at high frequencies.  
The discovery of a GPS radio source, considered to be a young object
where the radio emission started a few thousand years ago, but
displaying a steep spectral index ($\alpha_{\rm 1.6}^{8.4} > 1.0$)
in all its components is somewhat surprising. 
Indeed, the strong steepening found across the components of
PKS\,1518+047 is incompatible with the moderate high-frequency steepening
predicted by a continuous injection of particles, suggesting that no
injection/acceleration of fresh relativistic particles is currently
occurring in any region of the source.  
Furthermore, when the injection of fresh particles is over, the
strong adiabatic losses cause 
a fast shift 
of the spectral turnover towards lower frequencies and a decrement of the
peak flux density. Therefore, detections of dying radio sources with a
spectral peak still occurring in the GHz regime are extremely
rare. A possible explanation may arise from the presence of
a dense ambient medium enshrouding
the radio source which may confine the relativistic particles
reducing the adiabatic losses. This is probably the case of fading radio
sources found in galaxy clusters where the intracluster medium (ICM) 
is likely limiting
adiabatic cooling \citep{slee01}. However, in the case
of PKS\,1518+047 the source is surrounded 
by the interstellar medium (ISM)
of the host galaxy, but its confinement cannot be related to neutral
hydrogen, because observations searching for H$_{\rm I}$ absorption 
did not provide evidence of a dense environment \citep{gupta06}.\\
From the analysis of the optically-thin spectrum of PKS\,1518+047, we find
that the steepening may 
be explained assuming that the supply of relativistic plasma 
is still taking place, but the energy distribution of the injected particles is
uncommonly steep. 
Another explanation is that the steep spectral shape is due to
the absence of freshly injected/reaccelerated particles 
and the energy losses are driving the
spectrum evolution. Support to this scenario comes from the best fit
to the spectrum of the northern component, where the continuous
injection model fails in reproducing the spectral shape, even
assuming that relativistic particles are injected with a steep
energy distribution. This steep spectrum can be best reproduced
by a synchrotron model in which no particle supply is taking
place. Furthermore, the time spent by the source in its ``fader''
phase is about 20\% of the whole source age.\\
We estimate the break
energy $\gamma_{\rm b} $ of the electron spectrum and 
the electron radiative time by:  

\begin{equation}
\gamma_{\rm b} \sim 487 \nu_{\rm b}^{1/2} H^{-1/2} (1 + z)^{1/2}
\label{eq_gamma-nu}
\end{equation}

\noindent and

\begin{equation}
t_{s} = 5.03 \times 10^{4} H^{-3/2} \nu_{\rm b}^{-1/2} (1 +
z)^{-1/2} \,\;\;\; {\rm (yr)}
\label{age}
\end{equation}

\noindent where $H$ is in mG and $\nu_{\rm b}$ in GHz.\\ 
If in Eqs. \ref{eq_gamma-nu} and \ref{age} 
we consider the equipartition magnetic field
and $\nu_{\rm b}$ derived from the fits, 
we find that the energy of the electrons responsible for the break
should be $\gamma \sim$ 400 -- 600 and the source radiative lifetime
is $ t_{s} = 2700 \pm 600$ yr. 
As a consequence, electrons with $\gamma > 600$ (i.e. with shorter
radiative lifetime) have already depleted their energy and they
contribute only to the high-frequency tail of the spectrum.

\noindent If we consider the $t_{\rm OFF}/t_{s} \sim 0.2$, as derived from 
the fit to the overall spectrum, we find that $t_{\rm OFF} = 550 \pm 100$
years, and this represents the time elapsed since the last
supply/acceleration of relativistic particles. These values indicate
that the radio source was 2150$\pm$500 years old when the radio emission
switched off.\\
The small $t_{\rm OFF}$ derived suggests that adiabatic losses
had not enough time to shift the 
spectral peak of PKS\,1518+047 far from the GHz regime. In the presence of
adiabatic expansion, and considering a magnetic field frozen in the
plasma, the spectral peak is shifted towards lower frequencies:\\

\begin{equation}
\nu_{p,1} = \nu_{p,0} \left( \frac{t_{0}}{t_{1}} \right)^{4}
\label{peak}
\end{equation}

\noindent where $\nu_{p,0}$ and $\nu_{p,1}$ are the peak frequency at
the time $t_{0}$ and $t_{1}$ respectively \citep{mo08}. 
As the time
elapsed after the switch off increases, the peak moves
to lower and lower frequencies, making the source unrecognisable
as a young GPS. For example, when $t_{\rm OFF}$ will represent more
than 40\% of the total source lifetime, the peak
should be below 300 MHz, so it will be difficult to identify
as a dying young radio source. Future instruments
like LOFAR may unveil a population of such fading radio sources.\\

\section{Conclusions}

We presented results from multi-frequency VLBA and VLA observations 
of the GPS radio
source PKS\,1518+047. The analysis of the spectral index distribution 
across the whole source structure showed  
that all the source components have very steep optically-thin 
synchrotron spectra. The radio spectra are well explained by
  energy losses of relativistic particles after the
cessation of the injection of new plasma in the radio lobes. 
This result, together with the lack of the source core and active
non-steep spectrum components like hot spots and knots in the jets,
suggests that no injection/acceleration of fresh particle is
currently occurring in any region of the source. For this
reason, PKS\,1518+047 can be considered a fading young radio
source, in which the radio emission switched off shortly  
after its onset. 
When the supply of energy is switched off, given the
high magnetic fields in the plasma, the spectral turnover moves
rapidly towards low frequencies, making the source undetectable at the
frequencies commonly used for radio surveys. 
However, in PKS\,1518+047 the time elapsed
since the last particle acceleration is of the order of a few hundred
years, suggesting that
PKS\,1518+047 still has a GPS spectrum because adiabatic losses have not
had enough time to affect the source spectrum, shifting the peak
far from the GHz regime. If the interruption of the radio activity is
a temporary phase and the radio emission from the central engine
will re-start soon, it is possible that the source will appear again
as a GPS without the severe
steepening at high frequencies. If this does not happen, the fate of
this radio source is to emit at lower and lower frequencies, until it
disappears at frequencies well below the MHz regime.\\

\section*{Acknowledgemnet}

We thank the anonymous referee for carefully reading the manuscript and
valuable suggestions.
The VLBA is operated by the US National Radio Astronomy Observatory
which is a facility of the National Science Foundation operated under
a cooperative agreement by Associated University, Inc.
This work has made use of the
NASA/IPAC Extragalactic Database (NED), which is operated by the Jet
Propulsion Laboratory, California Institute of Technology, under
contract with the National Aeronautics and Space Administration.\\

\end{document}